\documentclass[prd,twocolumn,showpacs,preprintnumbers,nofootinbib,
superscriptaddress]{revtex4}
\usepackage{graphicx}
\usepackage{epsfig}
\usepackage{dcolumn}
\usepackage{bm}

\def\gsim{\lower0.5ex\hbox{$\:\buildrel >\over\sim\:$}}
\def\lsim{\lower0.5ex\hbox{$\:\buildrel <\over\sim\:$}}

\newcommand{\bea}{\begin{eqnarray}}
\newcommand{\eea}{\end{eqnarray}}

\def\d{\delta}

\begin{document}

\preprint{HIP-2004-19/TH}
\preprint{SLAC-PUB-10416}
\preprint{TIFR/TH/04-10}

\title{Higher Dimensional Models of Light Majorana Neutrinos  
Confronted by Data}

\author{JoAnne L. Hewett\footnote{Work supported in part by the
Department of Energy, Contract
DE-AC03-76SF00515}}%
\email{hewett@SLAC.Stanford.EDU}
\affiliation{Stanford Linear Acclerator Center, Menlo Park, CA 94025, USA}
\author{Probir Roy}%
\email{probir@theory.tifr.res.in}
\affiliation{Department of Theoretical Physics, Tata Institute of Fundamental 
Research, Homi Bhabha Road, Mumbai - 400 005, India}
\author{Sourov Roy}
\email{roy@pcu.helsinki.fi}
\affiliation{Helsinki Institute of Physics, P.O. Box 64, FIN-00014
University of Helsinki, Finland}

\received{\today}

\pacs{11.25.Mj, 14.60.Pq, 14.60.St} 

\begin{abstract} \vspace*{10pt}
We discuss experimental and observational constraints on certain 
models of higher dimensional light Majorana neutrinos. Models with 
flavor blind brane-bulk couplings plus three or four flavor diagonal 
light Majorana neutrinos on the brane, with subsequent mixing induced 
solely by the Kaluza-Klein tower of states, are found to be excluded by 
data on the oscillations of solar, atmospheric and reactor 
neutrinos, taken together with the WMAP upper bound on the sum of
neutrino masses. Extra dimensions, if relevant to neutrino mixing, 
need to discriminate between neutrino flavors. 
\end{abstract}

\maketitle

\noindent Do neutrinos have anything to do with large extra dimensions 
\cite{arkani-hamed}? Do these extra dimensions discriminate between
neutrino flavors? These questions have engaged the attention of many 
authors \cite{dienes} during the last few years. Let us consider the 
simple picture of a three-brane, containing the Standard Model (SM) fields, 
embedded in a higher dimensional space, known as the bulk. The basic 
idea then is to introduce one (or more) right chiral neutrino(s), 
which are singlet(s) with respect to the SM gauge group, in the 
bulk \cite{foot1}. Tiny physical masses for the active ${\rm SU(2)_L}$ 
doublet neutrinos on the brane, that mix among themselves as well as 
with sterile components, then get induced. These are controlled by the 
size(s) of the compactified extra dimension(s) and the strength(s) of the 
couplings(s) between the brane and the bulk states. The bulk states
expand into a Kaluza-Klein tower of states with a zero mode, as well as 
accompanying KK excitations of heavier sterile neutrinos with 
arithmetically progressing masses $M_k = \pm k/R$, k=1,2,... 
with ${\rm R}$ being the radius of the largest extra dimension. The 
leakage of a propagating active neutrino into these sterile states 
has to be kept controllably small so as not to significantly alter 
the conventional oscillation probability formula \cite{bilenky} in 
the standard picture of three directly mixed active neutrinos. An 
immediate concern, nonetheless, is whether this picture is compatible 
with the pattern of neutrino masses and mixing angles implied by the 
latest oscillation and cosmological data. This note will address 
precisely this issue, focussing on the question whether the 
brane-bulk couplings are flavor blind or whether they are
forced by all the data to have significant flavor dependence. 

A relevant basic issue is whether the physical massive neutrinos are 
Dirac or Majorana particles. There is a relatively straightforward picture 
in the former case. One combination of the bulk neutrinos now becomes 
the right chiral partner of an active neutrino on the brane. Together 
they acquire a Dirac mass $m_D \sim h v M_F/M_{Pl}$, where $v$ is 
the electroweak vacuum expectation value (VEV), $h$ is a Yukawa coupling 
strength (controlling the interaction between the brane and bulk states)
which is of order unity or less, $M_{Pl}$ the reduced Planck 
mass in (1+3) dimensions whereas $M_F$ is the higher dimensional 
fundamental scale, proposedly in the multi-TeV range. This scenario was  
given a thorough phenomenological examination in Ref. \cite{davoudiasl} 
with the conclusion that it is consistent with all extant experimental 
and observational constraints, provided $R^{-1} > 0.24$ eV. However, 
despite providing a natural explanation of the smallness of neutrino 
masses, this approach sheds no light on the observed pattern of 
neutrino mixing, nor on the hierarchy - if any - between the neutrino 
masses. These supposedly arise from the unknown structure of the 
Yukawa coupling matrix {\it h} in generation space. Any observed 
pattern of mixing and mass hierarchy among the neutrinos can be 
accommodated by this unknown structure, with the only constraint arising 
from the strict absence of neutrinoless double $\beta$-decay. A simple 
generalization of this picture with heavy right chiral Majorana 
neutrinos in the bulk leads to light physical Majorana neutrinos via 
the seesaw mechanism. There, one also needs to invoke (unknown) flavor 
dependent brane-bulk couplings.

There are higher dimensional scenarios \cite{dienes-sarcevic,lam-ng,
lam,cao} with a somewhat opposite approach. These assume tiny flavor
diagonal Majorana masses for left chiral neutrinos on the brane without 
explaining their smallness, but invoke extra dimensions to 
generate their mixing pattern. Here the higher dimensional KK modes 
induce a seesaw leading to light physical Majorana neutrinos with 
flavor mixing. However, this mixing pattern is strongly 
constrained. Suppose \cite{foot2} that (1) no flavor dependence is 
assumed in the couplings between the brane and bulk states and (2) 
flavor mixing is not allowed on the brane so that the brane Majorana 
neutrinos are kept flavor diagonal \cite{dienes-sarcevic,lam-ng,lam,cao}. 
Then, our observation is that the Dienes-Sarcevic model 
\cite{dienes-sarcevic,foot3}, with three flavor diagonal  
neutrinos on the brane, is ruled out by the data. The allowed 
parameter space actually gets quite constrained \cite{lam-ng} even 
when an arbitrary flavor dependence is included in the brane-bulk couplings. 
We also find that a version of the model, given in Ref. \cite{lam} with 
four flavor diagonal brane Majorana neutrinos and which has flavor blind 
brane-bulk couplings, is excluded. 

One attractive feature of the D--S model \cite{dienes-sarcevic} is 
the ability to provide a theory of neutrino mixing instead of just 
ascribing it to an unknown flavor dependence of the brane-bulk 
couplings. The intrinsic flavor diagonal Majorana masses on the 
brane for the active neutrinos $\nu_f$ ($f$ =1,2,3) are taken to 
be $m_f$, while the brane-bulk linkage is characterized by a 
universal mass $m$ appearing in all brane-bulk off-diagonal elements 
of the full neutrino Majorana mass matrix $\cal M$. Compactification 
to four dimensions yields the zero mode as well as to a tower of 
sterile neutrino KK states, as mentioned above. After the imposition
of appropriate orbifold conditions, the full neutrino Majorana mass 
matrix in the D-S model reads 
\begin{widetext}
\begin{eqnarray}
{\cal M} = \left(\matrix{m_1 & 0 & 0 & m & m & m & m & m & \ldots\cr
0 & m_2 & 0 & m & m & m & m & m & \ldots\cr
0 & 0 & m_3 & m & m & m & m & m & \ldots\cr
m & m & m & 0 & 0 & 0 & 0 & 0 & \ldots\cr
m & m & m & 0 & 1/R & 0 & 0 & 0 & \ldots\cr
m & m & m & 0 & 0 & -1/R & 0 & 0 & \ldots\cr
m & m & m & 0 & 0 & 0 & 2/R & 0 & \ldots\cr
m & m & m & 0 & 0 & 0 & 0 & -2/R & \ldots\cr
\vdots & \vdots & \vdots & \vdots & \vdots & \vdots & \vdots & 
\vdots & \ddots\cr }\right).
\label{massmatrix}
\end{eqnarray}
\end{widetext}
In consequence of (\ref{massmatrix}), the active neutrinos on the brane 
mix both among themselves and with the sterile KK states. Thus they undergo
flavor oscillations on propagation. Let $U$ be the unitary matrix 
which acts on mass eigenstate neutrino fields to yield flavor 
eigenstate neutrino fields. $U^T {\cal M} U$ is then diagonal. 
The matrix $U$ therefore is given by the inverse of the matrix of 
eigenvectors of $\cal M$ and is computable as such. The 
time-averaged probability $\overline {P_{f \rightarrow f^\prime}}$ 
that $|{\nu_f}\rangle$ oscillates into $|{\nu_{f^\prime}}\rangle$, 
with both $f$ and $f^\prime$ being flavor indices on the brane, is given 
by \cite{dienes,bilenky,dienes-sarcevic} 
\begin{eqnarray}
{\overline {P_{f \rightarrow f^\prime}}} = \d_{ff^\prime} - 2 \sum_{i=2}^N 
\sum_{j=1}^{i-1} {\rm Re} [U_{fi}U^*_{f^\prime i}U^*_{fj}U_{f^\prime j}],
\label{avprob}
\end{eqnarray}
\noindent where (\ref{avprob}), i and j are neutrino mass eigenstate
indices.

Let us now discuss the experimental results for the physical neutrino 
mass squared differences and the average neutrino oscillation 
probabilities. The atmospheric neutrino data yield \cite{foot5,superk}  
\begin{eqnarray}
\label{atmosmass}
1.3 \times 10^{-3}~{\rm eV}^2 & < & \Delta m^2_{\rm atm}  <  3.2 \times
10^{-3}~{\rm eV}^2, \\
0.9 & < & {\rm sin}^2{2\theta_{\rm atm}} \leq  1. 
\end{eqnarray}
\noindent The combined solar neutrino observations plus the 
KamLAND data imply \cite{sno,ahmed,garcia} 
\begin{eqnarray}
\label{solarmass}
6.1 \times 10^{-5}~{\rm eV}^2 & < & \Delta m^2_{\rm sol} < 9.0 
\times 10^{-5}~{\rm eV}^2, \\
0.71 & < & {\rm sin}^2{2\theta_{\rm sol}} < 0.91. 
\end{eqnarray}
\noindent The data \cite{chooz} from the CHOOZ reactor 
require that \cite{maltoni-valle}
\bea
{\rm sin}^2{2\theta_{13}} \leq 0.26. 
\label{choozdata}
\eea

\noindent In an effectively two flavor oscillation with a mixing angle 
$\phi$, the time-averaged oscillation probability is 
${\frac 1 2} {\rm sin}^2{2\phi}$. We can therefore use the estimates 
\bea
\label{atmosprob}
0.45 & < & \overline {P_{\rm atm}} < 0.5, \\
\label{solprob}
0.355 & < & \overline {P_{\rm sol}} < 0.455, \\
\label{choozprob}
\overline {P_{13}} & \leq & 0.13
\eea
\noindent in our analysis \cite{foot6}. 
Finally, the recent WMAP result puts \cite{spergel} an upper bound 
on the sum of the physical masses of stable light neutrinos that mix
among one another, namely 
\begin{eqnarray}
\sum_i m_i < 0.71~{\rm eV.}
\label{wmapconst}
\end{eqnarray}

\noindent A constraint, therefore, on any model is that there must 
be at least three light neutrinos with the sum of their masses 
not exceeding 0.71 eV.

It has been argued by Pierce and Murayama \cite{wmapmura} that the WMAP 
upper bound should include any light sterile neutrino that mixes with the 
active ones. This mixing is, of course, restricted to be rather small by 
the LEP constraint on invisible Z decay. A wide window for the existence
of a sterile neutrino is nonetheless allowed by the WMAP results. 
Suppose the active fraction of the sterile neutrino is $\epsilon$ 
so that $\epsilon^2 \sim {\rm sin}^{2}(2\theta_{mix})$, yielding  
$\epsilon \leq 10^{-1}$. Such a neutrino would have decoupled in the early 
universe at a temperature T where ${G_F^2}{T^5}{\epsilon^{2}} 
\sim {T^2}{M_{Pl}^{-1}}$. This temperature simply must be less than 
that of the QCD phase transition in order to avoid the dilution of 
those neutrinos by the entropy produced at this transition. Thus 
$\epsilon$ must lie in the the range $10^{-6} < \epsilon < 10^{-1}$. 
This means that in most regions of the parameter space of the models 
examined here, the summation in (\ref{wmapconst}) includes the 
contribution of light sterile neutrinos.

There are five parameters in the Dienes-Sarcevic (D-S) model : the input
Majorana masses $m_{1,2,3}$ for the three neutrinos on the brane, the universal
brane-bulk off-diagonal coupling $m$  and the radius of the compact extra
dimension $R$. For the present study we can set all mass scales in eV and
$R$ in ${\rm eV}^{-1}$. Let us also comment on the infinite number of sterile
neutrinos in the model. There are, of course, certain limiting regions of the
D-S parameter space when all the sterile neutrinos are sufficiently heavy
to be effectively decoupled from the phenomena leading to 
(\ref{atmosmass}) - (\ref{wmapconst}). There are other regions 
where one or more sterile states can become part of the light neutrino
spectrum. The existence of such light sterile neutrinos is already 
somewhat disfavored \cite{maltoni} by analyses of existing neutrino 
data. Though the situation is far from definite, results from the 
ongoing mini-Boone experiment are expected to provide a clearer conclusion.
At the moment, several light sterile neutrinos can still be accommodated.

We now come to a numerical investigation of
the D-S model in the five-dimensional parameter
space ($m_1$, $m_2$, $m_3$, $m$, $R^{-1}$). An initial
requirement is that there must be at least three
light neutrinos with the sum of their masses obeying
(\ref{wmapconst}). Thus those regions of the parameter space,
which are not consistent with this constraint, are immediately 
excluded. We attribute the masses $\mu_{1,2,3}$ to the physical 
light neutrinos, always ordered \cite{convention} such that 
$\mu_3 > \mu_2 > \mu_1$. The conditions (\ref{atmosmass}) and 
(\ref{solarmass}) then apply on the pertinent differences of squared 
mass eigenvalues. After employing these conditions, we have 
made a complete scan of the parameter space. Our quest is to 
determine if there is any region in which all the other constraints, 
i.e. (\ref{choozdata}) to (\ref{wmapconst}), can be satisfied. Our 
answer is in the negative, i.e. there is no point in the parameter 
space where all the four conditions can hold simultaneously. 
As will be shown below, there are parts of the parameter 
space where (\ref{atmosprob}), (\ref{choozprob}) and 
(\ref{wmapconst}) can be simultaneously satisfied. Similarly, there 
are other regions where the conditions of (\ref{solprob}), 
(\ref{choozprob}) and (\ref{wmapconst}) can be met. But these two sets 
of points do not overlap. Our conclusion is that the D-S model is ruled 
out by the data.

\begin{figure*}[htb]
\psfull
\begin{center}
\leavevmode
\epsfig{file=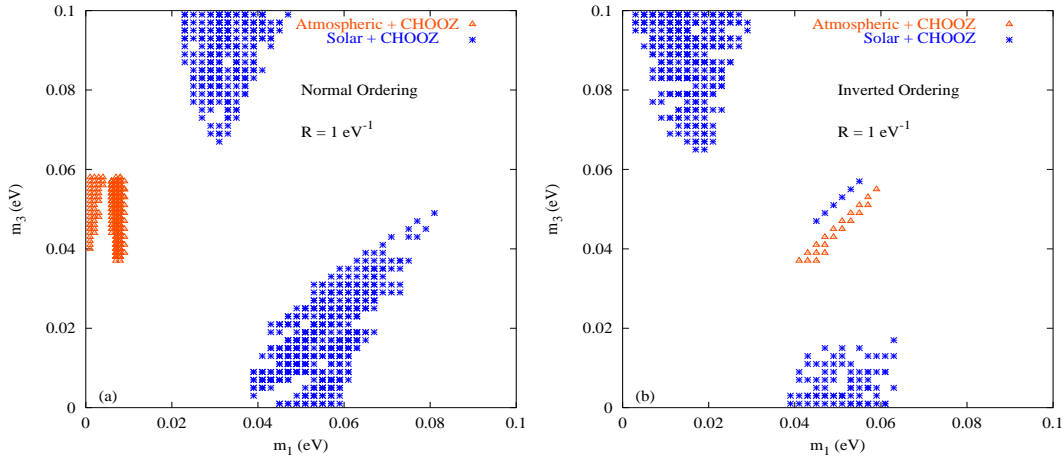,height=18cm,width=18cm,bbllx=0cm,bblly=2cm,
bburx=24cm,bbury=28cm,angle=0}
\end{center}
\vspace*{-4in}
\caption{\label{fig1} 
Allowed regions in the $m_1$ and $m_3$ plane for (a) the normal
ordering and (b) the inverted ordering of neutrino masses. Choices 
of parameters are described in the text.}
\end{figure*}

From practical considerations in making our scan, we have taken the 
number of contributing KK states N in (\ref{avprob}) to be finite 
though there actually are an infinite number of states in the KK tower. 
We choose N = 40 but have checked that {\it varying this truncation 
affects our results negligibly}. Furthermore, the choice of parameters 
in our numerical scanning was dictated by the eigenvalues of the mass-matrix 
(\ref{massmatrix}), as described below. Consider, for the moment, the 
possibility of  flavor dependent couplings for the brane neutrinos, 
leading to off-diagonal elements $m^\prime_i$ of $\cal M$. The 
eigenvalues of $\cal M$ are then determined by the characteristic equation
\bea
{1 \over \pi}{\tan(\pi\lambda)} = {\sum_{i=1}^3}{{m^{\prime2}_i R^2}\over
{\lambda-{m_i} R}}.
\label{character}
\eea 
Here, $\lambda = \mu R$ and $\mu$ represents the eigenvalues. Let us also 
define $d^2 \equiv R^2 ({\sum_{i=1}^3}{{m^\prime_i}^2})$ so that 
${m^\prime_i}R = de_i$, where $\sum_{i=1}^3{e_i^2 = 1}$. 
The two zeros of 
\bea
r(\lambda) \equiv {\sum_{i=1}^3}{{e_i^2}\over {\lambda-{m_i}R}}
\label{zeros}
\eea 
are given in the limit $m^\prime_1 = m^\prime_2 = m^\prime_3 = m$ (i.e., 
$e_1 = e_2 = e_3 = e$) by 
\bea
{\lambda_{b,c} \over R} &=& \mu_{b,c} = {1 \over 2} (p \mp c), \nonumber
                 \\[1.2ex]
p &=& 2{e^2}(m_1+m_2+m_3), \nonumber
                 \\[1.2ex]
c &=& {[p^2 - 4{e^2}(m_2 m_3 +m_3 m_1 + m_1 m_2)]}^{1/2}.
\label{lambdabc} 
\eea
It has been shown in Ref. \cite{lam-ng} that, as $d$ increases, two of the 
eigenvalues move towards $\mu_b$ and $\mu_c$ whereas the 
other eigenvalues move towards half-integral values times $R^{-1}$. 
For $d \gg 1$, the two eigenvalues sit at $\mu_b$ and $\mu_c$ 
(independent of $R$), while the rest are fixed at those half-integer 
values times $R^{-1}$. We have verified this property numerically. 
We have thus not included large values of $m~(mR = de)$ 
in our numerical scanning of the parameter space, since that would not 
allow for two different mass squared differences as required to explain 
both the solar and the atmospheric neutrino deficits. The other significant 
constraint on the range of $m_{1,2,3}$ is (\ref{wmapconst}). Let us 
now consider variations in the radius of compactification $R$. One 
should note first that, so long as the dimensionless products 
$m_1 R$, $m_2 R$, $m_3 R$ and $m R$ remain the same, one would get 
the same time-averaged oscillation probabilities. 
We have further checked that, for a fixed set of 
($m_1,m_2,m_3$ and $m$), our conclusions do not change with the 
variation in $R$ permitted by the constraints given in 
(\ref{atmosmass}), (\ref{solarmass}) and (\ref{wmapconst}). This 
way $R$ can be seen to be constrained in the range aproximately given by 
$10^{-6}~{\rm eV}^{-1} \lsim R \lsim 10~{\rm eV}^{-1}$. 

\begin{figure}
\includegraphics{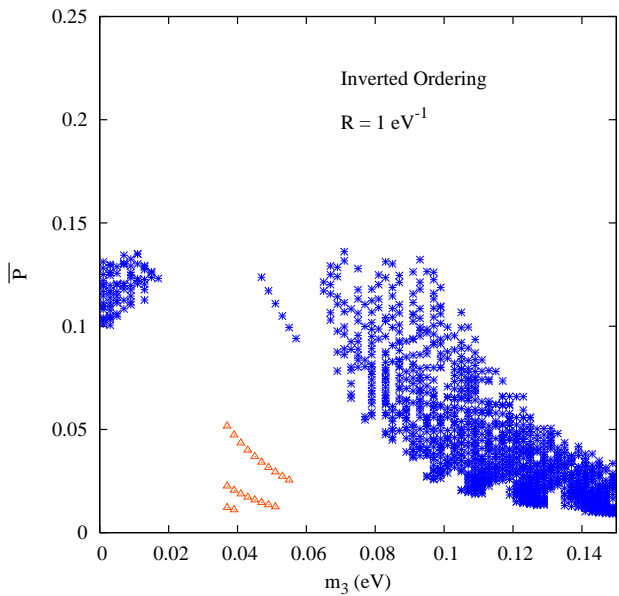}
\caption{\label{fig2}
Time-averaged oscillation probability (a) $\overline {P_{atm}}$ (in
Blue/star) and (b) $\overline {P_{sol}}$ (in orange/triangle) for 
the inverted ordering of neutrino masses as a function of $m_3$. Only 
those points are considered here which satisfy for (a) the Solar+CHOOZ 
and for (b) Atmospheric + CHOOZ constraints.}
\end{figure}
In our convention \cite{convention}, $\Delta m^2_{\rm sol} 
= \mu^2_2 - \mu^2_1$ and $\Delta m^2_{\rm atm} = \mu^2_3 - 
\mu^2_2$, for a normal ordering of neutrino masses, while 
$\Delta m^2_{\rm sol} = \mu^2_3 - \mu^2_2$ and 
$\Delta m^2_{\rm atm} = \mu^2_2 - \mu^2_1$ in the case of an inverted
ordering. Allowed regions, in the $m_1-m_3$ plane for the choice 
of $R^{-1}$ = 1 eV, are shown in Figs. \ref{fig1} (a) and (b).  The 
incompatibility of the D--S model with the presently known facts from
neutrino oscillation and cosmological data is made evident by these 
figures. In both figures $m_1$, $m_2$ and $m_3$ are taken to vary 
between 0.001 eV and 0.31 eV, whereas $m$ is varied between 
0.001 eV and 0.075 eV. Though, in the inverted case, a few points 
satisfying the conditions of (\ref{solprob}), (\ref{choozprob}) do 
lie close to the points satisfying (\ref{atmosprob}), 
(\ref{choozprob}), in fact there is not a single point in the 
five-dimensional parameter space of ($m_1, m_2, m_3, m, R^{-1}$) which 
is compatible with all the constraints. To further convince the 
reader, we display in Fig. \ref{fig2} the time-averaged oscillation 
probability (a) $\overline {P_{atm}}$ and (b) $\overline {P_{sol}}$ as 
a function of $m_3$ in the inverted case for all points shown in 
Fig.\ref{fig1}. Clearly, there is no consistent region of parameter
space. We have checked that this holds as R is varied in the range given
above.  

Next, we consider the model of Lam \cite{lam} which
requires a sterile neutrino (with a Majorana mass $m_4$) on the 
brane in addition to three active neutrinos. In the limit of large 
coupling between the brane neutrinos and the bulk (i.e., $d \gg {m_i}R$ 
,1) one can have three ``isolated" eigenvalues which can give rise 
to two different $\Delta m^2$, whereas the other eigenvalues sit at
half-integer values. Following Lam, we also arrange the Majorana masses
of the brane neutrinos in such a way that  
\bea
\label{mordering}
0 < m_1 < \mu_1 < m_2 < \mu_2 < m_3 < \mu_3 < m_4,
\eea 
\noindent where $\mu_i$ are the isolated eigenvalues.  
Furthermore, we assume that the ${m_i}R$ do not have  
integral or half-integral values. We take a flavor-blind 
brane-bulk coupling $m$ as in the earlier scenario. After a 
thourough scan in the parameter space with the above mentioned 
constraint in (\ref{mordering}), we find that although in some part 
of the parameter space it is possible to satisfy the constraints 
in (\ref{atmosmass}) and (\ref{solarmass}), there is no region 
where both the constraints in (\ref{atmosprob}) and (\ref{solprob}) 
can be satisfied either with a normal or with an inverted ordering of
the neutrino masses.   

In summary, we are able to exclude the Dienes-Sarcevic model
\cite{dienes-sarcevic} and that version of Lam's model \cite{lam} which
has flavor blind off-diagonal mass terms, by using all neutrino
oscillation data and the WMAP constraint on the sum of neutrino masses.
Extra dimensional models of neutrino masses and mixing angles need
flavor dependent brane-bulk couplings.

P.R. and S.R. acknowledge the hospitality of the Theory Group at SLAC
where this work was initiated. We thank S. Choubey, S. Goswami, A.S.
Joshipura and H-J. He for their helpful comments. The research of S.R.
was supported by the Academy of Finland (Project number 48787). S.R.
would also like to thank the Lady Davis Fellowship Trust in Technion,
Israel, for financial support during the initial stages of this work.


\begin{thebibliography}{99}
\bibitem{foot1}
There are more complicated
models consisting of right chiral neutrinos on the brane and/or left 
chiral ones in the bulk with warped geometric configurations 
\cite{grossman}, but the use of flavor dependent brane-bulk couplings 
to explain the observed pattern of neutrino mixing is quite generic in these.
\bibitem{foot2}
Thus the entire flavor mixing
among the light physical neutrinos is induced by the KK modes.
\bibitem{foot3}
An extension
of the model of Ref. \cite{dienes-sarcevic} with three different 
off-diagonal mass elements, was investigated in the limit of their 
very large values by Lam and Ng \cite{lam-ng} and shown to be 
incompatible with the data. Our study, however, covers the entire 
parameter space of the Dienes-Sarcevic model \cite{dienes-sarcevic} 
with a uniform off-diagonal mass.
\bibitem{foot6}
Our conclusions
follow just from these averaged probabilities without any need to
use the detailed shape of the corresponding distributions.
\bibitem{convention}
This convention of always keeping $\mu_3 > \mu_2 > \mu_1$, which is
different from the more common PMNS convention of attributing
$\mu_{1,2,3}$ to $|\nu_{e, \mu, \tau}\rangle$ in the `no-mixing' limit
$U \rightarrow I$, has been followed for search convenience and is
inessential to the derivation of the final conclusion.
\end{thebibliography}
\end{document}